\documentclass[prl,twocolumn,showpacs,amsmath,letter]{revtex4}
\usepackage{graphicx}

\newcommand{\Journal}[4]{#1 \textbf{#2}, #3 (#4)}

\begin{document}

\title{Current-Driven Magnetic Excitations in Permalloy-Based Multilayer Nanopillars}

\author{S. Urazhdin}
\author{Norman O. Birge}
\author{W. P. Pratt Jr.}
\author{J. Bass}
\affiliation{Department of Physics and Astronomy, Center for
Fundamental Materials Research and Center for Sensor Materials,
Michigan State University, East Lansing, MI 48824}

\pacs{73.40.-c, 75.60.Jk, 75.70.Cn}

\begin{abstract}
We study current-driven magnetization switching in nanofabricated
Ni$_{84}$Fe$_{16}$/Cu/Ni$_{84}$Fe$_{16}$ trilayers at 295~K and
4.2~K. The shape of the hysteretic switching diagram at low magnetic field changes with temperature.
The reversible behavior at higher field involves two phenomena, a threshold current for magnetic excitations closely
correlated with the switching current, and a peak in differential resistance characterized by telegraph noise,
with average period that decreases exponentially with current and shifts with temperature.
We interpret both static and dynamic results at 295~K and 4.2~K in
terms of thermal activation over a potential barrier, with a current dependent effective
magnetic temperature.
\end{abstract}

\maketitle

The effects of magnetic order on transport,
e.g. giant magnetoresistance, are well
studied and widely used in technology. The discovery of
predicted~\cite{slonczewski,berger} current-induced magnetization
precession~\cite{tsoiprl,tsoinature} and switching~\cite{cornellorig} stimulated
an explosion of interest in the reverse effect of the current on
magnetic order and magnetic dynamics. Besides fundamental interest
in the physics of magnetic systems driven far out of equilibrium,
these phenomena hold promise for
high-density memory applications. With the number of theoretical
papers growing rapidly~\cite{slonczewski2}-\cite{zhang2},
relatively few experimental facts are known, mostly for Co/Cu/Co
multilayers, and, for nanofabricated samples, mostly at room
temperature, T=295~K~\cite{cornellapl}-\cite{kent}.

By studying Py/Cu/Py (Py=Permalloy=Ni$_{84}$Fe$_{16}$)
nanofabricated trilayers (nanopillars), we are able for the first
time to quantitatively compare current-driven switching at 295~K
and 4.2~K. Much smaller crystalline anisotropy and magnetoelastic
coefficients of Py eliminate the irregular behavior at 4.2~K seen
in Co/Cu/Co by others~\cite{cornelltemp} and confirmed by us. Our
data let us study temperature dependences, and establish general
features of switching in both the hysteretic switching regime at
low magnetic field $H$ and the reversible switching regime at
higher $H$. We emphasize two important findings. (1) In a new
picture of the physics in the reversible regime, we isolate two
different phenomena: (a) magnetic excitations occuring above a
threshold current $I_t$, and appearing either as a linear rise or
peaks in differential resistance, $dV/dI$; and (b) the more well
known reversible switching peak. We show that the latter arises
from telegraph noise switching of magnetization between parallel
(P) and antiparallel (AP) orientations, and occurs when the dwell
times in the P and AP states are approximately equal,
$\tau_P\approx\tau_{AP}$. (2) Extending an idea of
Ref.~\cite{wegrowe}, we describe incoherent magnetic excitation by
current in both hysteretic and nonhysteretic switching regimes in
terms of an effective magnetic temperature $T_m$, which can differ
substantially from the lattice temperature $T_{ph}$. Our model
leads to a temperature dependence of the switching that differs
from expectation for models~\cite{cornelltemp,zhang2} based on
coherent current-driven excitation of the uniform
precession~\cite{slonczewski}.

Our samples were  nanofabricated with a
multistep process similar to that used by
others~\cite{cornellorig}. To minimize dipolar coupling between
the Py layers, we used the geometry of Albert {\it et
al.}~\cite{cornellapl}, with the bottom (thicker) Py layer
left extended, and the top Py layer and Cu spacer patterned
into an elongated shape with typical dimensions
$130\times 60$~nm. The bottom Py and the Cu spacer thicknesses
were always 20~nm and 10~nm, respectively (we give all thicknesses
in nm). The patterned Py layer thickness was varied from
2 to 6~nm. We measured differential resistances, dV/dI, with
four-probes and lock-in detection, adding an ac current of
amplitude 20~$\mu$A at 8~kHz to the dc current $I$. Most of 12
Py/Cu/Py devices studied had resistances $R\approx$1.5~$\Omega$,
magnetoresistance (MR) $\approx$5\% at 295~K, and MR$\approx$8\%
at 4.2~K. Positive current flows from the extended to the
patterned Py layer. $H$ is directed along the easy axis of the nanopillar.

\begin{figure}
\includegraphics[scale=0.4]{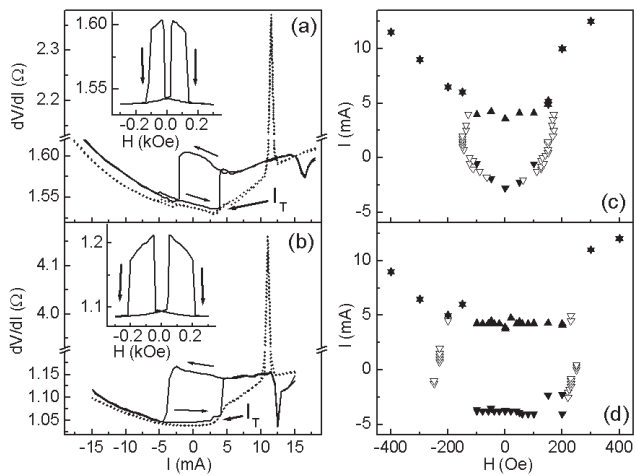}
\caption{\label{fig1} (a) Switching with current at 295~K in a
Py(20)/Cu(10)/Py(6) trilayer (thicknesses are in nm). Solid line: $H=50$~Oe, dashed line:
$H=-500$~Oe. Arrows mark the scan direction. $I_t$ is the threshold
current as defined in the text. Inset: MR dependence on $H$ at $I=0$.
(b)  Same as (a), at 4.2 K. (c),(d) Magnetization switching
diagram, extracted from the current-switching at fixed $H$ (solid
symbols), and field-switching at fixed $I$ (open symbols): (c) at 295
K, (d) at 4.2 K. Downward triangles: AP to P switching, upward: P
to AP switching. The reversible switching peaks are marked by
coinciding upward and downward triangles.}
\end{figure}

Figs.~\ref{fig1}(a,b) show the variations of $dV/dI$ with $I$ for
a patterned Py(20)/Cu(10)/Py(6) trilayer at 295~K (1a) and
4.2~K (1b), for $H=50$~Oe (solid curves) and $H=500$~Oe (dashed curves).
The insets show the variations with $H$ for $I=0$. At small $H$, the
magnetization switches hysteretically to a higher resistance
AP state at a large enough positive current
$I_s^{P\to AP}\equiv I_s$, and to a low resistance P state
at negative $I_s^{AP\to P}$. At larger $H$, the switching step
turns into a nonhysteretic peak.
Figs.~\ref{fig1}(c,d) show the switching diagrams at 295 K and 4.2 K,
extracted from data such as those in Figs.~\ref{fig1}(a),(b),
obtained both by varying $I$ at fixed $H$ and $H$ at fixed $I$.
Qualitatively, the 295~K data in Figs.~\ref{fig1}(a,c) are similar
to those published previously for Co/Cu/Co~\cite{cornellapl}. As
expected, both the reduced magnetization and
thermal activation result in smaller switching
currents and fields $H_s(I=0)$ at 295~K.  The
slight $H$-asymmetry in Figs.~\ref{fig1}(c,d) is attributed to a
combination of the current-induced Oersted field and sample shape
asymmetry.

\begin{figure}
\includegraphics[scale=0.4]{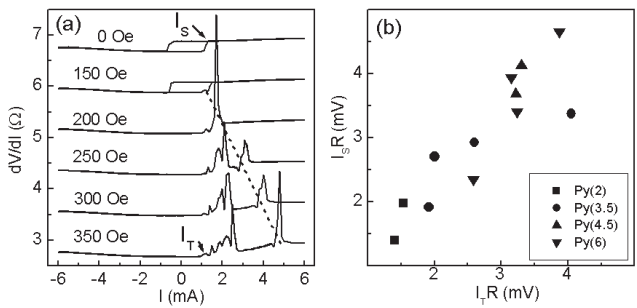}
\caption{\label{fig2}  (a) Differential resistance of a
Py(20)/Cu(10)/Py(3.5) trilayer for various $H$, at 4.2~K.
$R_P=3.6$~$\Omega$, $R_{AP}=4.0$~$\Omega$, curves are offset for
clarity. Dashed line follows the reversible switching peak. The
onset of the peaks in $R_P$ is marked by $I_t$, and $I_s$ is the
$P\to AP$ switching current at $H=0$.
(b) Relation between the $P\to AP$ switching current $I_s$ at $H=0$, $T=$4.2~K and the threshold
current $I_t$ (as defined in the text) for 12 samples with
patterned Py layer thicknesses: 2~nm (squares), 3.5~nm (diamonds), 4.5~nm (triangles), 6~nm (circles).
 $I_s$, $I_t$ are multiplied by the resistances $R$ of the samples.}
\end{figure}

The first important new feature of our data is the almost square
shape of the hysteretic region at 4.2~K (Fig.~\ref{fig1}(d)). At
4.2~K, the switching currents do not change in the range
$-230$~Oe$<H<230$~Oe, beyond which the switching becomes
reversible. Similarly, for $-4$~mA$<I<4$~mA, $H_s(I)$ is independent
of $I$, and at larger positive $I$ the switching becomes reversible.

The second important new feature is a threshold current $I_t$
(labeled in Figs.~\ref{fig1}(a,b) and Fig.~\ref{fig2}(a)) seen at
large $H$, with reversible switching. Most of our Py/Cu/Py
and Co/Cu/Co samples showed a nearly linear rise of $(dV/dI)_P\equiv R_P$ above $I_t$.
Similar behavior was likely present  in earlier Co/Cu/Co data at 295~K, but less obvious,
and thus rarely noted~\cite{grollier3}.
In our smallest sample, (Fig.~\ref{fig2}(a), dimensions estimated at $50\times100$~nm),
the linear rise is resolved at 4.2~K into several peaks. They appear only to the left of the reversible
switching peak, as the latter moves to higher $I$ with increasing $H$
(shown by a dashed line). We observed similar, less pronounced
peak structures in some samples with smaller $R$.
 Remarkably, in all
cases (e.g. Figs.~\ref{fig1}(a,b), Fig.~\ref{fig2}(a)), the onset
of a linear rise or peak structures in $R_P$ coincides with
$I_s(H=0)$. Within the uncertainty in determining the onset
current $I_t$, we find $I_s=I_t$ for all 12 samples studied
(Fig.~\ref{fig2}(b)). This correlation between $I_s$ and $I_t$,
combined with the abruptness of the changes with $I$ in
Fig.~\ref{fig1}(d), point to the importance of $I_t$.
Moreover, upon following the data up to $H=4$~kOe (not shown), we find $I_t$ to be
almost independent of $H$ (increasing by only 10--40\% for different samples).

\begin{figure}
\includegraphics[scale=0.44]{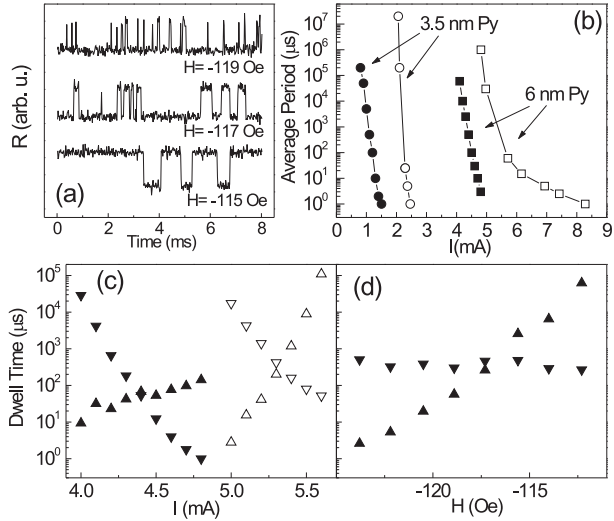}
\caption{\label{fig3} (a) Time traces of the Py(20)/Cu(10)/Py(6)
sample static resistance $R=V/I$ at $I=4.4$~mA, $T=$295~K at different
$H$. (b) Current dependence of the average telegraph noise period
for Py(20)/Cu(10)/Py(6) (squares) and Py(20)/Cu(10)/Py(3.5)
(circles) samples. Open symbols: $T=$4.2~K, filled symbols: $T=$295~K.
$H$ was adjusted so that the average dwell times in AP and P states
were equal. (c) The dependence of the P (downward triangles) and
AP (upward triangles) dwell times on $I$, for a Py(20)/Cu(10)/Py(6)
sample. Open symbols: $T=$4.2~K, $H=-335$~Oe, filled symbols: $T=$295~K,
$H=-120$~Oe. (d) Dependence of P (downward triangles) and AP
(upward triangles) dwell times on $H$ at $I=4.4$~mA, $T=295$~K.}
\end{figure}

We identify $I_t$ as a threshold for magnetic excitation of the
patterned Py layer. An excitation threshold was previously seen as a peak in the differential resistance
of point-contacts on extended magnetic multilayers~\cite{tsoiprl,tsoinature}. In nanopillars,
this threshold behavior is modified into a linear rise or peak
structures. The reversible switching peak appears at higher $H$, and evolves faster with $H$
(dashed line in Fig.~\ref{fig2}(a)) than the peaks we attribute to magnetic excitations.
Time-resolved measurements of resistance, at $I$ and $H$ close to
the reversible switching peak (Fig.~\ref{fig3}(a)), show that the reversible switching is characterized
by telegraph noise with random distribution of $\tau_P$,
$\tau_{AP}$. Similar slow telegraph noise was reported in a point
contact at one current~\cite{tsoiprl}, and at the transition point
from hysteretic to reversible switching in Co/Cu nanopillars at
295~K, at $I$ or $H$ fixed~\cite{cornelltemp}.

The data of Fig.~\ref{fig3}(b,c), and the finding that the
reversible switching peak is a consequence of telegraph noise
and occurs at $\tau_P\approx \tau_{AP}$,
represent our third new experimental result. Fig.~\ref{fig3}(b)
shows that, when both $I$ and $H$ are increased so as to hold
$\tau_P=\tau_{AP}$, the average period of the telegraph noise
decreases exponentially with similar slopes at 295~K and 4.2~K,
down to the 1~MHz bandwidth limit of our setup. Fig.~\ref{fig3}(c)
shows that the variations of $\tau_P,\tau_{AP}$ with $I$ have
similar forms at 295~K and 4.2~K. These data (as well as similar
data for Co/Cu at 295~K~\cite{cornelltemp,unpublished}) show that
the reversible switching peak does not indicate abrupt onset of a
new physical process. At small $I$, $\tau_P>>\tau_{AP}$, and the
average sample resistance is close to $R_P$, while at large $I$,
$\tau_P<<\tau_{AP}$, and the average resistance is close to
$R_{AP}$. The reversible switching peak appears in $dV/dI$ at
$\tau_P\approx \tau_{AP}$, due to the exponential variation of
$\tau_P,\tau_{AP}$ with $I$. Thus, it merely reflects the
current-dependent telegraph noise statistics; $I_t$ is the only
fundamental current related to magnetic excitations. The amplitude
and inverse width of the reversible switching peak are
proportional to $d[(\tau_{AP}-\tau_P)/(\tau_{AP}+\tau_P)]/dI$. In
Fig.~\ref{fig2}(a), they are correlated with the positions of the
peaks that we attribute to magnetic excitations. For example, at
200~Oe, the reversible peak is on top of the magnetic excitation
peaks, and is significantly taller and narrower than at higher
$H$. More detailed data and analysis will be presented
elsewhere~\cite{tobepublished}.

We interpret both our static and dynamic results consistently in
terms of a model involving thermally activated switching over a
one-dimensional potential barrier, with an effective
current-dependent temperature $T_m(I)$~\cite{wegrowe}, and a
threshold current, $I_t$. $T_m$ can differ from the lattice
temperature $T_{ph}$ in confined geometry, where the magnetic
energy relaxes significantly more slowly than the highly excited
individual magnetic modes~\cite{book}. Formally, a potential
energy barrier is inconsistent with the current-driven torque in
the Landau-Lifshitz-Gilbert equation~\cite{slonczewski}. Li and
Zhang~\cite{zhang2} have shown that the work performed by the
torque leads to a current-dependent effective switching barrier,
but with $T_m=T_{ph}$, which we shall see is inconsistent with our
data of Fig.~\ref{fig3}(b,c).

Our observation that $I_t$ varies only weakly with $H$ up to 4~kOe
provides information about whether its origin is quantum
mechanical~\cite{berger} or classical~\cite{slonczewski}. In the
quantum model, $I_t$ follows from $\Delta\mu=\hbar\omega$, where
$\Delta\mu$ is the current dependent difference in chemical
potentials of spin-up and spin-down electrons due to spin
accumulation, and $\omega$ is the magnon frequency.
At small $H$, $I_t\propto\omega\propto\sqrt{1+H/H_a}$~\cite{kittel},
where $H_a$ is the in-plane anisotropy
field. In our samples, $H_a\approx 300$~Oe, giving much too rapid an increase of $I_t$ with $H$.
In the classical treatment based on the Landau-Lifshitz-Gilbert equation, $I_t$ is set
by the balance between the torque induced by the spin polarized
current and Gilbert damping. The $H$ dependence of $I_t$ is much
weaker $I_t(H)\approx I_t(0)[1+H/(2\pi M)]$~\cite{cornellorig},
where $M$ is the magnetization, $M\approx 880$~Oe for Py. This
model predicts $\frac{dI_t(H)}{I_t(0)dH}\approx 0.2$~(kOe)$^{-1}$, not too far
from the measured values of 0.03--0.1 for different samples.

\begin{figure}
\includegraphics[scale=0.35]{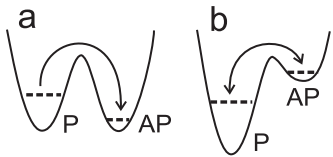}
\caption{\label{fig4} (a) Schematic of current-driven hysteretic
magnetization switching at $H=0$. Dashed lines indicate $T_m$. (b)
Schematic of telegraph noise at $H>H_s$.}
\end{figure}

To analyze our data quantitatively, we determine the dwell times $\tau_{P,AP}$ by
\begin{equation}\label{expdec}
\tau_{P,AP}=\frac{1}{\Omega}exp\left[\frac{U_{P,AP}}{kT^{P,AP}_m}\right],
\end{equation}
and approximate $T_m$ by the heuristic relation
\begin{equation}\label{tm}
T_m=T_{ph}+K(I-I_t)\mbox{ for }I>I_t.
\end{equation}
Here $K$ is a constant, $\Omega\approx 10^7s^{-1}$~\cite{koch} is the effective
attempt frequency, $U_{P,AP}$ is the potential barrier height for
switching from the P or AP state, and $T_m^{AP}(I)\not= T_m^P(I)$ when $I\not= 0$.
The switching barriers $U_{P,AP}$ depend on $I$
only through the variation of the magnetization with temperature
$T_m$. In the P state, at $I>I_t$, magnetic excitation leads to
increase of $T_m^P$, as illustrated in Fig.~\ref{fig4}(a). A
thermally activated transition into the AP state occurs at
$kT_m^P\approx\frac{U_P}{ln(t_{exp}\Omega)}\approx
\frac{U_P}{16}$, based on the data acquisition time $t_{exp}$ of 1
second per point. The conditions for magnetic excitation are not
satisfied in the AP state, so the magnetic system cools to
$T_m^{AP}\approx T_{ph}$ and at $H<H_s$ becomes trapped in this
state. At $H>H_s$ (Fig.~\ref{fig4}(b)),
$kT_{ph}>\frac{U_{AP}}{16}$, i.e. the $AP\to P$ transition is also
thermally activated, leading to telegraph noise at $I>I_s$,
$H>H_s$.

The 4.2~K switching diagram of Fig.~\ref{fig1}(d)) is consistent
with this model. It reflects the weak variations of $I_t$ with
$H$, and of $T_m$ (and thus $H_s$) with $I$ below $I_t$. We
attribute the rounding of the 295~K diagram at $I<0$ to
enhancement of thermal fluctuations of magnetization by current,
not included in the heuristic Eq.~(\ref{tm}). Increasing H
decreases $U_{AP}$, so that even these weak excitations can
activate the AP$\to$P transition. This process
is described in different terms in Ref.~\cite{zhang2}.

The dependencies of nonhysteretic switching on $I$, $H$, and $T$
shown in Fig.~\ref{fig3}(b-d) agree with
Eqs.~(\ref{expdec}) and (\ref{tm}). As $I$ increases with $H$ fixed, $\tau_P$
decreases exponentially because of the increase of $T_m^P$ with
$I>I_t$. An increase of $\tau_{AP}$ with $I$ may be evidence for
cooling of the magnetic system of the patterned layer in the AP
state at $I>0$, through a magnon absorption mechanism inverse to
the magnon emission in the P state. At $T_{ph}$=4.2~K, the requirement
for telegraph noise ($U_{AP}<16kT_{ph}$) is satisfied with much
smaller $U_{AP}$ than at 295~K. Hence a slight decrease in
$T_m^{AP}$ with $I$ has a stronger effect on $\tau_{AP}$ at 4.2~K
than at 295~K, as seen in Fig.~\ref{fig3}(c). The dependencies in
Fig.~\ref{fig3}(d) follow from $\frac{\partial
ln\tau_{P,AP}}{\partial H}=\frac{1}{kT_m^{P,AP}}\frac{\partial
U_{P,AP}}{\partial H}$. Since $T_m^P>T_m^{AP}$, $\tau_{P}$ varies
with $H$ more slowly than $\tau_{AP}$. Based on the data of
Fig.~\ref{fig3}, we estimate the dependence $T_m(I)$ numerically.
Some of our Py(20)/Cu(10)/Py(3.5) and all of the
Py(20)/Cu(10)/Py(2) samples were superparamagnetic at 295~K.
Assuming linear variation of the switching barrier with nanopillar
thickness, we estimate the switching barrier $U_{P,AP}(H=0)\approx
0.7$~eV for Py(20)/Cu(10)/Py(6) samples. We use an approximate
dependence $U(T_m)=U_0\sqrt{1-T_m/T_c}$, where $T_c=800$~K is the Curie
temperature for Py. Using
$d\ln(\tau_P)/dI=11$~(mA)$^{-1}$ from Fig.~\ref{fig3}(c), we obtain from
Eq.~(\ref{expdec}) $K=\frac{dT_m^P}{dI}\approx 400$~K/mA.
Both the 295~K (filled symbols in Fig.~\ref{fig3}(c)) and 4.2~K data (open symbols)
are approximately consistent with  this estimate.  In contrast, starting from a current-dependent effective barrier with
$T_m=T_{ph}$~\cite{cornelltemp, zhang2}, one would predict
$\frac{d\ln (\tau_P(I))}{dI}\approx \frac{U_P}{I_ckT_{ph}}$, where
$I_c$ is the switching current at $T=0$. This strong dependence on $T_{ph}$
disagrees with the similarity of the 4.2~K and 295~K data.

To summarize, our major new experimental results on Py/Cu/Py
nanopillars are: i) a square switching diagram at 4.2~K; ii) an
onset current $I_t$ (closely related to the hysteretic switching current $I_s$)
for a linear rise of $dV/dI$ in larger samples or a series of peaks in smaller ones; iii)
reversible switching, characterized by telegraph noise
with rate both increasing exponentially with $I$ and shifting with temperature.
The reversible switching peak in $dV/dI$ occurs when the dwell times in the P and AP
states are approximately equal.
We are able to explain the hysteretic switching behavior at both 295~K and 4.2K,
and variations of telegraph noise with $I$, $H$, and $T$, by means
of a threshold current for the onset of magnetic excitations and
thermally activated switching with a current-dependent magnetic temperature.

We acknowledge helpful communications with R. Loloee, H. Kurt, M.I. Dykman, M.D. Stiles,
A.H. Macdonald, D.C. Ralph, S. Zhang, A. Fert, support from the MSU CFMR, CSM, the MSU
Keck Microfabrication facility, the NSF through Grants DMR
02-02476, 98-09688, and 00-98803, and Seagate Technology.


\begin{thebibliography}{99}
\bibitem{slonczewski} J. Slonczewski, \Journal{J. Magn. Magn. Mater.}{159}{L1}{1996}.
\bibitem{berger} L. Berger, \Journal{Phys. Rev.}{B 54}{9353}{1996}.
\bibitem{tsoiprl} M. Tsoi {\it et al.}, \Journal{Phys. Rev.
Lett.}{80}{4281}{1998}; \textbf{81}, 493(E) (1998).
\bibitem{tsoinature} M. Tsoi {\it t al.}, \Journal{Nature (London)}{406}{46}{2000}.
\bibitem{cornellorig} J. A. Katine {\it et al.}, \Journal{Phys. Rev. Lett.}{84}{3149}{2000}.
\bibitem{slonczewski2} J. Slonczewski, \Journal{J. Magn. Magn. Mater.}{195}{L261}{1999}.
\bibitem{fmrdc} L. Berger, \Journal{Phys. Rev.}{B 59}{11465}{1999}.
\bibitem{waintal} X. Waintal {\it et al.}, \Journal{Phys. Rev.}{B 62}{12317}{2000}.
\bibitem{sun} J. Z. Sun, \Journal{Phys. Rev.}{B 62}{570}{2000}.
\bibitem{brataas} A. Brataas, Yu. V. Nazarov, and Gerrit E. W. Bauer, \Journal{Phys. Rev. Lett.}{84}{2481}{2000}.
\bibitem{heide} C. Heide, P. E. Zilberman, and R. J. Elliott, \Journal{Phys. Rev.}{B 63}{064424}{2001}.
\bibitem{predict} Ya. B. Bazaliy, B. A. Jones, and S. C. Zhang, \Journal{J. Appl. Phys.}{89}{6793}{2001}.
\bibitem{zhang} S. Zhang, P. M. Levy, A. Fert, \Journal{Phys. Rev. Lett.}{88}{236601-1}{2002}.
\bibitem{stiles} M. D. Stiles and A. Zangwill, \Journal{Phys. Rev.}{B 66}{014407-1}{2002}.
\bibitem{zhang2} Z. Li and S. Zhang,  cond-mat/0302339.
\bibitem{cornellapl} F. J. Albert {\it et al.}, \Journal{Appl. Phys. Lett.}{77}{3809}{2000}.
\bibitem{grollier} J. Grollier {\it et al.}, \Journal{Appl. Phys. Lett.}{78}{3663}{2001}.
\bibitem{grollier2} J. Grollier {\it et al.}, \Journal{Phys. Rev.}{B 67}{174402}{2003}.
\bibitem{cornelltemp} E. B. Myers {\it et al.}, \Journal{Phys. Rev. Lett.}{89}{196801}{2002}.
\bibitem{cornellquant} F. J. Albert {\it et al.}, \Journal{Phys. Rev. Lett.}{89}{226802}{2002}.
\bibitem{wegrowe} J. E. Wegrowe {\it et al.}, \Journal{J. Appl. Phys.}{91}{6806}{2002}.
\bibitem{sun2} J. Z. Sun {\it et al.}, J. Appl. Phys. (in press).
\bibitem{kent} B. Oezyilmaz {\it et al.}, cond-matt/0301324.
\bibitem{unpublished} S. Urazhdin {\it et al.}, cond-mat/0303614.
\bibitem{grollier3} In Ref.~\cite{grollier2}, a rise of the P state resistance was noted, but described as a
progressive reversal to the AP state, that would not display the widely seen reversible switching peak in $dV/dI$.
\bibitem{tobepublished} S. Urazhdin {\it et al.}, to be published.
\bibitem{book} J. Miltat, G. Albuquerque, and A. Thiaville in {\it Spin Dynamics in Confined Magnetic Structures I}, Springer, New York, 2002.
\bibitem{kittel} The field dependence of the uniform magnon frequency is given in Eq.(41) Ch.(16) of C. Kittel, {\it Introduction to Solid State Physics}, 7th edition, John Wiley \& Sons, NY, 1996.
\bibitem{koch} R. H. Koch {\it et al.}, \Journal{Phys. Rev. Lett.}{84}{5419}{2000}.
\end{thebibliography}
\end{document}